\documentclass[conference,a4paper]{APSIPA2021}
\usepackage{multirow}
\usepackage[dvips]{graphicx}
\usepackage{amsmath}
\usepackage{amsxtra}
\usepackage{threeparttable}

\usepackage{cite} 
\usepackage{hyperref}
\usepackage{graphicx}
\usepackage{float}

\usepackage{CJKutf8}

\usepackage{geometry}
\usepackage{fancyhdr}

\fancypagestyle{firststyle} {
    \fancyhf{}
    \fancyhead[L]{Proceedings of 2022 APSIPA Annual Summit and Conference}
    \fancyhead[R]{7-10 November 2022, Chiang Mai, Thailand}
    
    \fancyfoot[L]{978-616-590-477-3 ©2022 APSIPA}
    \fancyfoot[C]{\thepage}
    \fancyfoot[R]{APSIPA ASC 2022}
}
\fancypagestyle{fancy} {
    \fancyhf{}
    \fancyhead[L]{Proceedings of 2022 APSIPA Annual Summit and Conference}
    \fancyhead[R]{7-10 November 2022, Chiang Mai, Thailand}
    
    \fancyfoot[C]{\thepage}
}

\begin{document}
\begin{CJK*}{UTF8}{gbsn}
\title{Back-Translation-Style Data Augmentation for Mandarin Chinese Polyphone Disambiguation}

\author{%
\authorblockN{%
Chunyu Qiang$^*$, Peng Yang$^*$, Hao Che, Jinba Xiao, Xiaorui Wang, Zhongyuan Wang
}
\authorblockA{%
\authorrefmark{0}
Kwai, Beijing, P.R. China \\
E-mail: \{qiangchunyu, yangpeng, chehao, xiaojinba, wangxiaorui, wangzhongyuan\}@kuaishou.com}
}
\maketitle

\begin{abstract}
Conversion of Chinese Grapheme-to-Phoneme (G2P) plays an important role in Mandarin Chinese Text-To-Speech (TTS) systems, where one of the biggest challenges is the task of polyphone disambiguation. Most of the previous polyphone disambiguation models are trained on manually annotated datasets, and publicly available datasets for polyphone disambiguation are scarce. In this paper we propose a simple back-translation-style data augmentation method for mandarin Chinese polyphone disambiguation, utilizing a large amount of unlabeled text data. Inspired by the back-translation technique proposed in the field of machine translation, we build a Grapheme-to-Phoneme (G2P) model to predict the pronunciation of polyphonic character, and a Phoneme-to-Grapheme (P2G) model to predict pronunciation into text. Meanwhile, a window-based matching strategy and a multi-model scoring strategy are proposed to judge the correctness of the pseudo-label. We design a data balance strategy to improve the accuracy of some typical polyphonic characters in the training set with imbalanced distribution or data scarcity. The experimental result shows the effectiveness of the proposed back-translation-style data augmentation method.
\end{abstract}

\renewcommand{\thefootnote}{\fnsymbol{footnote}} 
\footnotetext[1]{These authors contributed equally to this work.} 
 
\section{Introduction}
With the development of deep learning, speech synthesis technology has rapidly advanced
\cite{sotelo2017char2wav, wang2017tacotron,peng2020non,kim2020glow,liu2021vara,elias2021parallel}. 
In TTS system, the front-end text processing module significantly influences the intelligibility and naturalness of synthesized speech. G2P is an essential component in the front-end module of the TTS systems, which one of the biggest challenges is how to disambiguate the pronunciation of polyphones characters. 

Traditional approaches for polyphone disambiguation are rule-based algorithms\cite{zhang2002efficient, huang2008disambiguating} and statistical machine learning methods\cite{zhang2001disambiguation, mao2007inequality, liu2010polyphonic, liu2011polyphone, shan2016bi,cai2019polyphone}.  The rule-based approach chooses the pronunciation of the polyphonic character based on predefined complex rules along with a dictionary. However, this requires a substantial amount of linguistic knowledge. In contrast, the data driven approach employ statistical methods such as Decision trees (DT) or Maximum Entropy (ME) \cite{zhang2001disambiguation, mao2007inequality, liu2010polyphonic, liu2011polyphone}. The recent tremendous success of the neural network in various fields has prompted polyphone disambiguation to turn to neural network-based models. Shan et al.\cite{shan2016bi} and K.Park et al.\cite{park2020g2pm} adopted bidirectional long-short-term memory (BLSTM) layer to predict the pronunciation of polyphonic character that is in accord with the context. Dai et al. \cite{dai2019disambiguation} proposed a method combining the pre-trained BERT \cite{devlin2018bert} with a neural-network based classifier. Cai et al.\cite{cai2019polyphone} proposed a polyphone disambiguation systems with multi-level conditions respectively. Zhang et al.\cite{zhang2020distant} constructed a framework based on CNN in a distantly supervised way. Zhang et al. proposed a mask-based model \cite{zhang2020mask} and a system using FLAT \cite{zhang2021polyphone} for polyphone disambiguation. Li et al.\cite{li2021improving} proposed a novel system based on word-level features and window-based attention for polyphone disambiguation. These advancements are mainly contributed by the application of supervised learning to expensive datasets. Nonetheless, publicly available datasets for polyphone disambiguation are scarce\cite{park2020g2pm, shi2021polyphone}. At the same time, there are few studies on data augmentation for Mandarin Chinese polyphone disambiguation\cite{shi2021polyphone, zhang2022data}.

Back-translation is a method which has been proposed in the field of machine translation \cite{sennrich2015improving, lample2017unsupervised}. In this approach, a pre-trained target-to-source translation model is used to generate source text from the unpaired target text. Augmenting training data with back-translated data led to notable improvement in performance of neural machine translation models. Inspired by the back-translation approach, in this paper we propose a simple data augmentation method for mandarin Chinese polyphone disambiguation. We build a G2P model to predict the pronunciation of polyphonic character, and a P2G model used to predict pronunciation into text. Meanwhile, we propose two pseudo-label screening strategies. Window-based matching strategy is to use fixed-length windows on both sides of polyphonic words for text matching to judge the correctness of pseudo-labels. The multi-model scoring strategy uses multiple polyphone models to predict the pronunciation to judge the correctness of the pseudo-label. After training, the G2P model generates the pronunciation labels from a large number of unpaired texts, the P2G model predicts the original text from the pronunciation labels, the screening strategies select the available data, and the G2P model is retrained using the generated pronunciations as additional training data. To the best of our knowledge, we are the first to apply back-translation-style data augmentation on the polyphone disambiguation task to reduce the dependence on expensive manually labeled data.

\begin{figure*}[t]
 \centering
 \includegraphics[width=\textwidth]{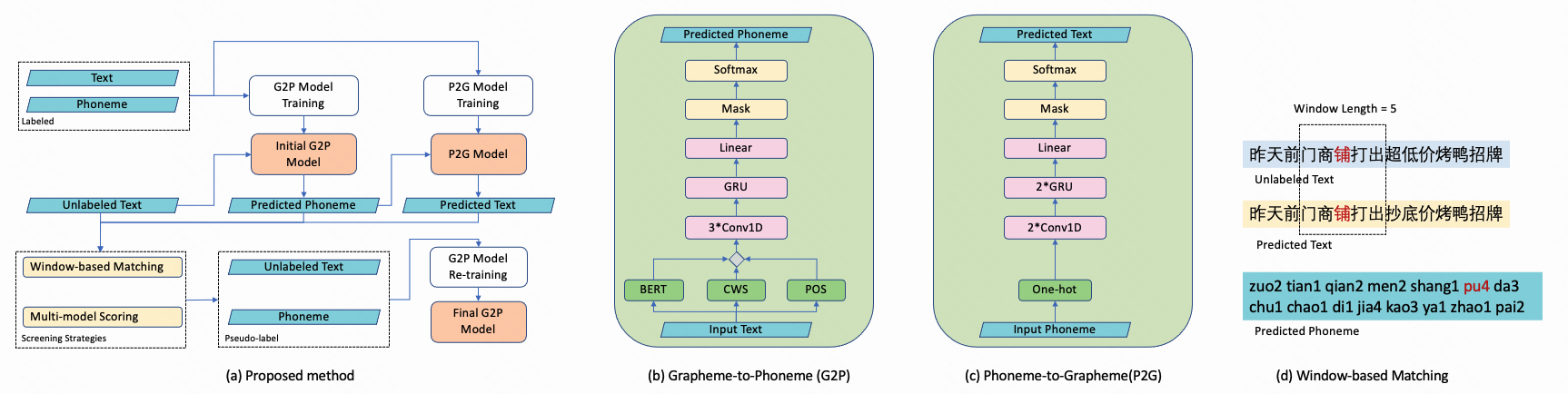}
 \caption{The architecture of (a) Proposed method, (b) G2P, (c) P2G, (d) Window-based Matching.}
 \label{fig:speech_production}
\end{figure*}


The contributions of this paper are as follows.

\begin{itemize}
\item A simple back-translation-style data augmentation method is proposed for mandarin Chinese polyphone disambiguation.
\end{itemize}

\begin{itemize}
\item A window-based matching strategy and a multi-model scoring strategy are proposed to judge the correctness of the pseudo-label. 
\end{itemize}

\begin{itemize}
\item A data-balanced method is designed to address the uneven distribution of characters and pronunciations in polyphones.
\end{itemize}

\section{Method}

\subsection{Overview}
An overview of our proposed method is shown in Figure 1 (a). We build a G2P model to predict the pronunciation of polyphonic character, and a P2G model to predict pronunciation into text. The two models are trained using paired training data which consists of text and prounciation. After training, the G2P model generates the pronunciation labels from a large number of unpaired texts, and the P2G model predicts the original text from the pronunciation labels. Then, the screening strategies selects the available data and retrains the G2P model using the generated pronunciations as additional training data.


\subsection{Grapheme-to-Phoneme}
The conventional polyphone disambiguation serves to convert the input polyphonic characters into their corresponding pinyin. We regard the polyphone disambiguation as a classification task. Recently, connecting the existed networks with a pre-trained BERT is the de facto way of improving performance in the field of NLP. As illustrated in Figure1 (b), we propose a mask-based model and a system by using BERT for G2P. According to existing work, Chinese word segmentation (CWS) and part of speech (POS) tagging are essential to the mandarin Chinese polyphone disambiguation task. Therefore, we apply the Chinese characters of the input sentence and the corresponding lexical information as input features, such as CWS and POS tagging. Firstly, a character sequence $X = [x_1,x_2,...,x_n]$  is fed into the pre-trained BERT encoder to get the character-level hidden features: $H = [h_1,h_2,...,h_n]$. Secondly, we use an external CWS module to segment $X$ into a sequence of word: $W = [w_1,w_2,...,w_m] (m \leq n)$. Then, we use a POS module for part-of-speech tagging to get a sequence of part-of-speech: $P = [p_1, p_2,...,p_m]$. The length of the word sequence $W$ and part-of-speech sequence $P$ is the same as the length of the word sequence $X$ by the method of up-sampling. Sequence $W$, sequence $P$ and sequence $H$ are concatenated and input into prediction network. We connect three sequential 1D-convolutional layers with a GRU layer and a fully-connected linear layer as the prediction network. Furthermore, we ignore all impossible prounciations that do not belong to the target character in ways that add a mask layer that sets the score of the corresponding classes to -inf before the final softmax layer. 


\subsection{Phoneme-to-Grapheme}
P2G can be used for precdicting the corresponding character sequence from the input pinyin sequence. Similar to the G2P task, we treat P2G as a classification task. As illustrated in Figure1 (c), we propose a mask-based model for P2G. The pinyin sequence is encoded by one-hot method and input to the prediction network. We connect two sequential 1D-convolutional layers with two GRU layers and a fully-connected linear layer as the prediction network. Furthermore, we ignore all impossible characters that do not belong to the target prounciation in ways that add a mask layer that sets the score of the corresponding classes to -inf before the final softmax layer.


\subsection{Pseudo Labels}
The key to semi-supervised learning for polyphone disambiguation task is to assign pseudo-labels for unlabeled data. This paper proposes two pseudo-label screening strategies, including window-based matching and multi-model scoring.
\subsubsection{Window-based Matching}

Window-based matching strategy is to use fixed-length windows on both sides of polyphonic characters for text matching to judge the correctness of pseudo-labels.  As illustrated in Figure1 (d), an example of an unlabeled text "昨天前门商铺打出超低价烤鸭招牌" (Yesterday, the Qianmen shop displayed the signboard of the ultra-low price roast duck),which the corresponding pinyin sequence "zuo2 tian1 qian2 men2 shang1 pu4 da3 chu1 chao1 di1 jia4 kao3 ya1 zhao1 pai2" is predicted through the G2P model. And then, through the P2G model, the pinyin sequence gets the predicted text "昨天前门商铺打出抄底价烤鸭招牌" (Yesterday, the Qianmen shop displayed the signboard of the bottom price roast duck). In this sentence, only "铺" is a polyphonic word that needs to be predicted, while other words can directly get the corresponding pronunciation by looking up the table.  When the window length is 5, the string on both sides of "铺" is "门商铺打出", the unlabeled text is consistent with the predicted text, and the pseudo-label is considered to be available. However, when the window length is 7, the unlabeled text is inconsistent with the predicted text, and the pseudo-label is considered unavailable ("前门商铺打出超" vs "前门商铺打出抄"). In this method, the window length is a key parameter that can apparently affects the accuracy.

\begin{table}[b]\centering
\caption{Statistical information of corpus}
\begin{tabular}{|c|c|c|c|c|}
\hline
Character & \begin{tabular}[c]{@{}c@{}}Character \\ Count\end{tabular} & \begin{tabular}[c]{@{}c@{}}Character \\ Frequency\end{tabular} & Pinyin                                                                    & \begin{tabular}[c]{@{}c@{}}High-fre\\ Pinyin Frequency\end{tabular} \\ \hline
的         & 111550                                                     & 3.36\%                                                    & \begin{tabular}[c]{@{}c@{}}de5: 111126\\ di4: 297\\ di2: 130\end{tabular} & 99.62\%                                                        \\ \hline
了         & 30941                                                      & 0.93\%                                                    & \begin{tabular}[c]{@{}c@{}}le5: 30514\\ liao3: 428\end{tabular}           & 98.62\%                                                        \\ \hline
地         & 14141                                                      & 0.43\%                                                    & \begin{tabular}[c]{@{}c@{}}di4: 9608\\ de5: 4534\end{tabular}             & 67.94\%                                                        \\ \hline
重         & 5716                                                       & 0.17\%                                                    & \begin{tabular}[c]{@{}c@{}}chong2: 1404\\ zhong4: 4313\end{tabular}       & 75.44\%                                                        \\ \hline
背         & 650                                                        & 0.02\%                                                    & \begin{tabular}[c]{@{}c@{}}bei4: 517\\ bei1: 134\end{tabular}             & 79.42\%                                                        \\ \hline
帖         & 54                                                         & 0.00\%                                                    & \begin{tabular}[c]{@{}c@{}}tie3: 37\\ tie4: 14\\ tie1:4\end{tabular}      & 66.07\%                                                        \\ \hline
\end{tabular}
\end{table}

\begin{table}[b]\centering
\caption{The accuracy of different system}
\resizebox{!}{!}{
\begin{tabular}{|c|c|}
\hline
System                                       & Test Accuracy  \\ \hline
G2Pm (BLSTM) \cite{park2020g2pm}                        & 87.88\%          \\ \hline
MASK-BASED  \cite{zhang2020mask}                        & 88.32\%          \\ \hline
G2Pm (BERT)  \cite{park2020g2pm}                         & 89.23\%          \\ \hline
Base                                         & 88.92\%          \\ \hline
BTSDA (window=1)                              & 88.17\%          \\ \hline
BTSDA (window=3)                              & 90.87\%          \\ \hline
BTSDA (window=5)                              & 91.38\%          \\ \hline
BTSDA (window=7)                              & 91.30\%           \\ \hline
BTSDA (window=max)                            & 90.28\%          \\ \hline
\textbf{BTSDA (window=5)+Multi-model-scoring} & \textbf{91.47\%} \\ \hline
\end{tabular}}
\end{table}

\begin{table}[b]\centering
\caption{The accuracy of polyphonic characters}
\begin{tabular}{|c|c|c|c|c|}
\hline
Character & \begin{tabular}[c]{@{}c@{}}Polyphone \end{tabular} & \begin{tabular}[c]{@{}c@{}}Polyphone\\ (Data Balance)\end{tabular} & Base                                                                    & \begin{tabular}[c]{@{}c@{}}BTSDA\end{tabular} \\ \hline
重         & \begin{tabular}[c]{@{}l@{}}chong2 : 1404\\ zhong4 : 4313\end{tabular} & \begin{tabular}[c]{@{}l@{}}chong2 : 2475\\ zhong4 : 4458\end{tabular} & 47.46\% & 80.60\% \\ \hline
相         & \begin{tabular}[c]{@{}l@{}}xiang1 : 3254\\ xiang4 : 463\end{tabular}  & \begin{tabular}[c]{@{}l@{}}xiang1 : 4699\\ xiang4 : 1307\end{tabular} & 53.13\% & 71.31\% \\ \hline
夏         & \begin{tabular}[c]{@{}l@{}}sha4 : 169\\ xia4 : 171\end{tabular}       & \begin{tabular}[c]{@{}l@{}}sha4 : 1636\\ xia4 : 1642\end{tabular}       & 50.29\% & 74.19\% \\ \hline
晕         & \begin{tabular}[c]{@{}l@{}}yun1 : 69\\ yun4 : 121\end{tabular}        & \begin{tabular}[c]{@{}l@{}}yun1 : 1494\\ yun4 : 1632\end{tabular}       & 63.68\% & 93.40\%  \\ \hline

\end{tabular}
\end{table}

\subsubsection{Multi-model Scoring}
The multi-model scoring strategy uses multiple polyphone models to predict the pronunciation with the aim of judging the correctness of the pseudo-label. We construct multiple G2P models by modifying the prediction network structure, and predict the pronunciation of polyphonic words for unlabeled texts by these models. If the results of all models are consistent, the pseudo-label is considered to be available and effective.

\subsection{Data Balance}
Table I illustrates the statistical information of corpus. In polyphone corpus, there are common problems of uneven distribution of characters and uneven distribution of pronunciations, attaching excessive attention to massive and easily classified examples resulting the model less precise in terms of rare and hard classified examples, thereby degrading the performance of the system. In response to the above problems, we adopt a data balance method.
\subsubsection{Character Data Balance}
As shown in Table I, the frequency of common characters (such as "的", "了") in the corpus is much higher than that of uncommon characters (such as "背"，“帖”). Using the data augmentation method proposed in this paper, the frequency of all polyphonic characters is not less than 0.1\%. In this way, character data balance enables the model to better classify rare and hard characters.
\subsubsection{Pronunciation Data Balance}
As shown in Table I, the frequency of the common pronunciation ("de5") of "的" in the corpus is much higher than that of the uncommon pronunciation ("di2"). Using the data augmentation method proposed in this paper, 
the frequency of each pronunciation is not less than  20\% in the corresponding polyphonic character. In this way, pronunciation data balance enables the model to better classify rare and hard pronunciations.

\section{Experiments}

\subsection{Dataset}
We use a human-annotated Chinese polyphonic character dataset in our experiments. There are total of 859,112 sentences in the dataset and each sentence contains one or more polyphonic character and its corresponding correct pinyin annotation. While the training set contains 687,290 sentences, the dev set contains 85,911 sentences and the test set contains 85,911 sentences, and the phoneme of each polyphonic character is annotated by native Mandarin Chinese speakers. The dataset of which the length of ranges from 3 to 350, contains 650 polyphonic characters in total with corresponding 334 pinyins. As for unlabeled data, we collect 2,209,271 lines of text from Internet.

\subsection{Experimental Step}
cThe details of the experimental systems are listed as follow:

\begin{itemize}
 \item [1]
\textbf{G2Pm (BLSTM)}: Same as the structure described in \cite{park2020g2pm}. The system does not use any external language processing tools such as word segmenter, entity recognizer, or part-of-speech tagger. Instead, the system takes as input a sequence of characters and train the network in the end-to-end manner. The system adopts BLSTM layer to predict the pronunciation of polyphonic character that is in accord with the context. 
\end{itemize}

\begin{itemize}
 \item [2]
\textbf{MASK-BASED}: Same as the structure described in \cite{zhang2020mask}. We implemented a MASK-BASED based model for comparison. We use a word segmentation model based on CNN, GRU and CRF to obtain CWS and POS features. The MASK-BASED model consists of BLSTM and 1D-CNN structures. The weight-softmax mechanism is used in the experiment, and Modified Focal Loss is selected as the loss function.
\end{itemize}

\begin{itemize}
 \item [3]
\textbf{G2Pm (BERT)}: Same as the structure described in \cite{park2020g2pm}. The system attachs a fully connected layer to the BERT network and feed the hidden state of the polyphonic character to it. 
\end{itemize}

\begin{itemize}
 \item [4]
\textbf{Base}: As illustrated in Figure1 (b), we propose a mask-based model and a system by using BERT for G2P. We apply the Chinese characters of the input sentence and the corresponding lexical information, such as CWS and POS tagging, as input features. The system does not use the back-translation-style data augmentation (BTSDA) method proposed in this paper
\end{itemize}

\begin{itemize}
 \item [5]
\textbf{BTSDA (window=1)} : Same as system 4 but applied back-translation-style data augmentation method in the model additionally.  At the same time, the pseudo-label screening strategy of window-based matching is used, and the window length parameter is set to 1.
\end{itemize}

\begin{itemize}
 \item [6]
\textbf{BTSDA (window=3)} : Same as system 5 but the window length parameter is set to 3.
\end{itemize}

\begin{itemize}
 \item [7]
\textbf{BTSDA (window=5)}: Same as system 5 but the window length parameter is set to 5.
\end{itemize}

\begin{itemize}
 \item [8]
\textbf{BTSDA (window=7)}: Same as system 5 but the window length parameter is set to 7.
\end{itemize}

\begin{itemize}
 \item [9]
\textbf{BTSDA (window=max)}: Same as system 5 but the window length parameter is set to the maximum length of the input sentence.
\end{itemize}

\begin{itemize}
 \item [10]
\textbf{BTSDA (window=5)+Multi-model-scoring}: Same as system 7 but applied the pseudp-label screening strategy of multi-model-scoring.
\end{itemize}

\subsection{Results and Analysis}
In this section, we compare the method proposed in this paper with the existing polyphonic disambiguatio models. For the fairness of the comparison, all models are trained on the same dataset. As shown in Table II, the base model proposed in this paper outperforms the G2Pm (BLSTM) and MASK-BASED models, but slightly worse than the G2Pm (BERT) model. Since base model is based on the pre-trained language model and lexical information, it has significantly improved compared with the G2Pm (BLSTM) and MASK-BASED model. We can also observe that the back-translation-style data augmentation method (BTSDA) significantly improves the performance compared to base model. The BTSDA method is used to match pseudo-labels for a large amount of unlabeled text. We also increase the window length from 1 to 7 so as to analyze the influence of window length. Table II shows that when only matching the current polyphone character (window=1), the accuracy is lower than other conditions, which is caused by the addition of incorrect pseudo-label data. The best performance is achieved when matching the context of two characters (window=5), but results degrades when expanding window length to the maximum length of the input sentence. This means that an excessively long window length will filter out more correct pseudo-label data. We demonstrate the effectiveness of multi-model scoring strategy on System 10, which achieves the best performance with a 2.55\% improvement in accuracy relative to the base model. It means that multi-model scoring strategy can improve the screening accuracy of pseudo-labels.

As shown in Table III, we compare the BTSDA (window=5 + Multi-model-scoring) model with the base model to demonstrate the effectiveness of the data balance method on some typical polyphonic characters suffering from imbalanced distribution or data scarcity in the training set. For those characters with serious imbalance problems such as "重" and "相", and characters with insufficient data such as "夏" and "晕", BTSDA model gets higher predicting accuracy compared with base model. "重" and "相" are augmented by the pronunciation data balance strategy, and the accuracy is increased by 33.14\% and 18.18\% respectively. "夏" and "晕" are augmented by the character data balance strategy, and the accuracy is increased by 23.90\% and 29.72\% respectively. The back-translation-style data augmentation method proposed in this paper makes the model more robust and generalizable.

\section{Conclusions}

In this paper we propose a simple back-translation-style data augmentation method for mandarin Chinese polyphone disambiguation, utilizing a large amount of unlabeled text data. Inspired by the back-translation technique proposed in the field of machine translation, we build a G2P model to predict the pronunciation of polyphonic character, and a P2G model to predict pronunciation into text. Meanwhile, a window-based matching strategy and a multi-model scoring strategy are proposed to judge the correctness of the pseudo-label. In addition, we design a data balance strategy to improve the accuracy of some typical polyphonic characters in the training set with imbalanced distribution or data scarcity. These experimental results indicate the effectiveness of the proposed back-translation-style data augmentation method and generate correct pseudo-labels on a large amount of unsupervised data, which greatly enhances the performance of polyphone disambiguation. In future works, we will optimize the model structure of G2P and P2G to improve the prediction accuracy, and apply our proposed method using much larger amount of unpaired text.

\newpage

\bibliographystyle{IEEEtran}



\end{CJK*}
\end{document}